\documentclass[12pt]{spieman}  
\usepackage{amsmath,amsfonts,amssymb}
\usepackage{graphicx}
\usepackage{setspace}
\usepackage{tocloft}
\usepackage[normalem]{ulem}
\usepackage{soul}

\title{Initial Fabrication and Characterization of Chemically-Etched Silicon Slits for KOSMOS}

\author[a]{Debby Tran}
\author[a,b,*]{Sarah Tuttle}
\author[a,b,c]{Kal Kadlec}
\author[a,b, d]{Rishi Pahuja}
\author[a]{Ali C. Jones}
\author[b]{William Ketzeback}
\author[b]{Russet McMillan}
\author[b]{Amanda Townsend}
\affil[a]{University of Washington, Astronomy Department, Seattle, 98195, USA}
\affil[b]{Apache Point Observatory, Sunspot, NM, 88349}
\affil[c]{Las Cumbres Observatory, 6740 Cortona Drive, Suite 102, Goleta, CA, 93117, USA}
\affil[d]{Caltech Optical Observatories, Pasadena, CA, USA}

\cftpagenumbersoff{figure}
\cftpagenumbersoff{table} 
\begin{document} 
\maketitle

\begin{abstract}
KOSMOS is a low-resolution, long-slit, optical spectrograph that has been upgraded at the University of Washington for its move from Kitt Peak National Observatory's Mayall 4m telescope to the Apache Point Observatory's ARC 3.5m telescope. One of the additions to KOSMOS is a slitviewer, which requires the fabrication of reflective slits, as KOSMOS previously used matte slits machined via wire EDM. We explore a novel method of slit fabrication using nanofabrication methods and compare the slit edge roughness, width uniformity, and the resulting scattering of the new fabricated slits to the original slits. We find the kerf surface of the chemically-etched reflective silicon slits are generally smoother than the machined matte slits, with an upper limit average roughness of 0.42 $\pm$ 0.03 $\mu$m versus 1.06 $\pm$ 0.04 $\mu$m respectively. The etched slits have width standard deviations of 6 $\pm$ 3 $\mu$m versus 10 $\pm$ 6 $\mu$m, respectively. The scattering for the chemically-etched slits is higher than that of the machined slits, showing that the reflectivity is the major contributor to scattering, not the roughness. This scattering, however, can be effectively reduced to zero with proper background subtraction. As slit widths increase, scattering increases for both types of slits, as expected. Future work will consist of testing and comparing the throughput and spectrophotometric data quality of these nanofabricated slits to the machined slits with on-sky data, in addition to making the etched slits more robust against breakage and finalizing the slit manufacturing process.
\end{abstract}


{\noindent \footnotesize\textbf{*}Sarah Tuttle,  \linkable{tuttlese@uw.edu} }

\begin{spacing}{2}   

\section{Introduction}\label{sect:intro}  
KOSMOS (the KPNO (Kitt Peak National Observatory) Ohio State Multi-Object Spectrograph) is a low-resolution (R $\sim$ 2100) long-slit optical spectrograph covering wavelength ranges of approximately 350 nm - 1 $\mu$m \cite{10.1117/12.2056834}. KOSMOS utilizes red and blue volume phase holographic (VPH) gratings to optimize throughput in the optical bandpass. It is one of a pair of twin instruments commissioned in late 2013/early 2014 - COSMOS at Cerro Tololo Inter-American Observatory (CTIO) and KOSMOS at KPNO. They are modified versions of OSMOS (Ohio State Multi-Object Spectrograph) currently in use on MDM (Michigan-Dartmouth-MIT) Observatory's 2.4m Hiltner telescope \cite{2011PASP..123..187M}. In April 2018, the Dark Energy Survey Instrument (DESI) became the primary instrument for the 4m Mayall telescope at KPNO. The installation of DESI required the telescope to be taken apart and reconfigured to accommodate this new instrument, retiring KOSMOS and other 4m instruments. 

KOSMOS is now seeing second light at Apache Point Observatory's (APO) ARC 3.5m telescope. A general purpose low-resolution optical spectrograph is an important tool for science targets that may be faint or not well-characterized. One of the initial instruments commissioned for the ARC 3.5m was the Dual Imaging Spectrograph\cite{DIS} (DIS), which has been heavily used as the low resolution (R $\sim$ 1000-7000) optical spectrograph at APO since the early 1990s. Now that it is approximately 30 years old, it is currently struggling with some deterioration of performance as a function of age. KOSMOS updates the telescope's optical spectroscopic capability, after being modified for the APO observing community. 

Additions to KOSMOS include a slitviewer camera, internal calibration lamps, a second internal electronics box to support the aforementioned modifications, and a Nasymth Port (NA2) adapter (Kadlec et al, in prep). These features will improve the usability of KOSMOS, particularly as APO has shifted over time to enhance capabilities to support time domain follow up and other time-specified observations. Having internal calibration lamps, for example, allows observers to run calibrations at the pointing for improved wavelength calibration, or outside of their observing windows while the instrument is off the telescope. The slitviewer is essential for ensuring the target is on the slit without using valuable observing time to switch between imaging and spectroscopic modes. Along with its use as a target acquisition camera, the slitviewer integrates with the field and boresight guiding software to function as an on-axis guider. Furthermore, the long slit on KOSMOS will allow users to take spectra of multiple aligned objects simultaneously.

Of the many modifications made to KOSMOS, this paper focuses on the requirement of a library of reflective slit masks. Prior to the addition of the slitviewer, KOSMOS utilized matte, machined slits to minimize scattering. Reflective slits redirect the photons that would have been scattered by matte slits towards the slitviewer camera, giving users the ability to know where they are in the sky. With new slits required for the retrofit of KOSMOS, we had the opportunity to test and document an alternative way of manufacturing slits - via wet chemical-etching, which is the preferred method of selective material removal in the semiconductor industry. As such, the paper adopts the same technology and investigates the feasibility of this method of fabrication for slits. The process outlined in this paper was the pilot  manufacturing for KOSMOS's reflective slits and will be improved upon in future iterations of slits. In the process of testing these new slits, we were able compare the slit edge roughness and uniformity of the slits fabricated via chemical-etching and wire EDM (electrical discharge machining). This allows us to test whether the roughness of the slit is an important factor in achieving higher spectrophotometric precision and throughput.

\subsection{Motivation}
\begin{figure}[h] 
    \centering
    \includegraphics{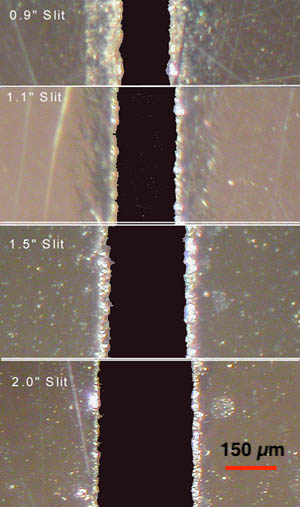}
    \caption{Example reflective slits fabricated via wire EDM for DIS at the ARC 3.5m \cite{DIS}}
    \label{fig:dis-slits}
\end{figure}

Wire EDM is a common method of fabricating spectrograph slits. This is an electrothermal production process where a thin single strand metal wire cuts through metal by the use of heat from electrical sparks. This method has been one of the most consistent, precise, and cost effective methods of machining slits \cite{headland_2020}. Despite this, there is still room for improving the consistency and uniformity of these slits. The slit edge roughness of the wire EDM slits from DIS (Figure \ref{fig:dis-slits}) is visually apparent, especially upon comparison to the chemically-etched slits in Figure \ref{fig:remaining-nitride}. Typical roughness as a result of wire EDM is on the order of micron scales \cite{JOSHI2017158}, while typical roughness from wet-etching is on the scale of nanometers \cite{Ezoe:06}. With nanofabrication facilities available at many research universities (including the University of Washington), we wanted to test the feasibility of the semiconductor device fabrication methods for producing ultra-flat reflective slits as demonstrated in Ref. \citenum{10.1117/12.672369} and compare their uniformity and performance. Many of these fabrication processes and their benefits and constraints as applied to astronomical instruments are not well-documented. As these techniques have become more widely available, in this paper, we attempt to better constrain requirements, make measurements, and document their impact on astronomical data.

Below we present the manufacturing and characterization of chemically-etched reflective slits, a required upgrade to accommodate the new slitviewing capability of KOSMOS at APO. In Section \ref{sect:Design}, we discuss the design approach for the new slits. In Section \ref{sect:manufacturing}, we discuss the manufacturing process for the nanofabricated slits. In Section \ref{sect:Data}, we discuss the acquisition and processing of data taken via microscope and in lab with KOSMOS. In Section \ref{sect:Discussion}, we present the slit edge roughness and width results of the matte and reflective slits and their impact on the data.

\section{Design}\label{sect:Design}
\begin{figure}[h] 
    \centering
    \includegraphics[width=\textwidth]{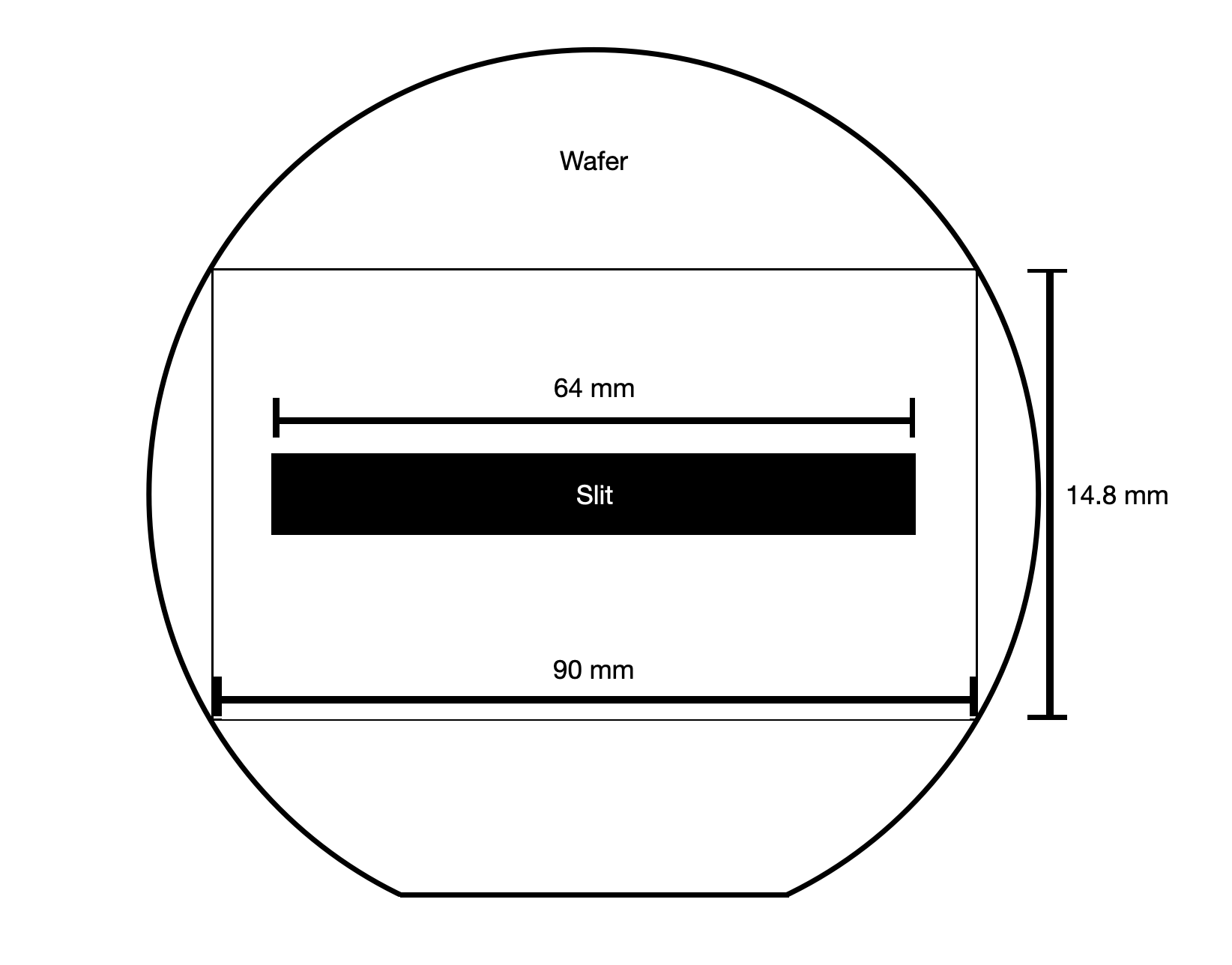}
    \caption{The thin lines of the rectangle outline the dicing lanes where the wafer is diced. The slit length is indicated, but slit width is not as the width varies depending on design specifications. For this work, achieved slit widths were 0.36" to 7.3" (64 $\mu$m to 1250 $\mu$m). The image is not to scale.}
    \label{fig:slit-mask-dim}
\end{figure}

Until recently, the functional field of view (FOV) at APO was 6.1 arcminutes. Recent upgrades to the baffles increased this to 8 arcminutes. To align with these upgrades and acquire data across the entire FOV, we attempted to make the slits 80mm long (8.6 arcminutes). However, this made the wafers fragile. The stress of its own weight would break the wafers along the slits while sitting in their carriers in between fabrication steps. Of the 75 slits fabricated, all 25 8.6 arcminute-long wafers broke during various steps of the manufacturing process. Each 525 $\mu$m thick, 100 mm long wafer had a remaining 5 mm on each side of the slit, which was simply not enough to support the the sizable through-wafer hole in the center. Unfortunately, dicing the wafer to reduce strain prior to the etching was not an option due to constraints of the tools. To create more stability, we shortened the length of the slits to be the length of the original KOSMOS slits, 64 mm long (6.8 arcmins; Figure \ref{fig:slit-mask-dim}). This is consistent with the size limitations due to the current shutter 90mm in diameter (Kadlec et al., in prep).

Selection of desired slit widths goes from 0.5" to 20" (29 $\mu$m to 1169.6 $\mu$m), with these limits chosen based on being seeing-limited at 0.5" and 20" for a user-specific project. However, due to an error with the plate scales of the instrument and telescope, the achieved widths have become 0.36" to 7.3" (64 $\mu$m to 1250 $\mu$m); Table \ref{tab:ref-width}). Since the matte slits from the previous iteration of KOSMOS are from 0.58" to 2.92" (91 $\mu$m to 456 $\mu$m); Table \ref{tab:matte-width}), this selection of reflective slit widths still expands upon the previously available slit widths for KOSMOS.

\section{Manufacturing}
\label{sect:manufacturing}
The reflective slits for KOSMOS are fabricated using photolithography, wet chemical etching, and vapor deposition, which is a typical process for fabricating semiconductor devices. Based on Atalla's work on surface passivation by thermal oxidation \cite{Atalla1960}, Hoerni patented the planar process in 1959 \cite{Hoerni1959}. The planar process is a manufacturing process and serves as the basis for small-scale and mass production of integrated circuits in the semiconductor industry, as well as the groundwork for this project. It is composed of photolithography, wet-chemical development, and metal deposition. To optimize for mass production, the industry uses standard silicon wafers and various applications of the planar process. This process is widely-used for mechanical and electrical silicon structures, such as membranes, nozzles, diaphragms, and trenches, due to the nanometer-scale precision that photolithography allows in patterning \cite{Micromachining_transducers,1456599}. Because this process is standard, nanofabrication facilities are widely available at the university level. 

\begin{figure}
\begin{center}
\begin{tabular}{c}
\includegraphics[scale=.5]{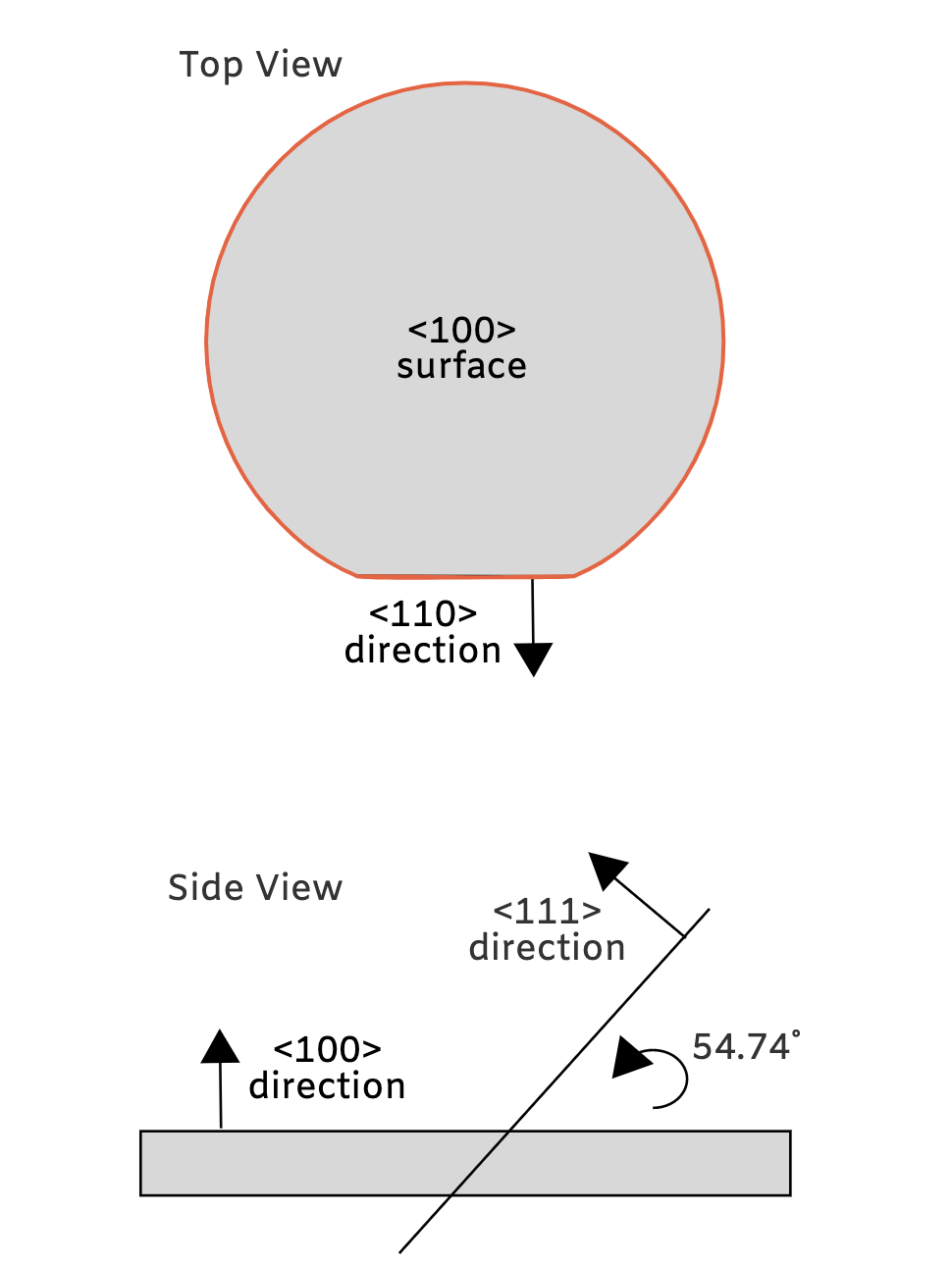}
\end{tabular}
\end{center}
\caption 
{ \label{fig:wafer}
Top and side view of $\langle 100 \rangle$ silicon wafer, where the wafer surface is a $\langle 100 \rangle$ plane with the wafer flat along the $\langle 110 \rangle$ plane. The $\langle 111 \rangle$ plane is 54.74$^{\circ}$ from the $\langle 100 \rangle$ plane. All planes are orthogonal to one another.} 
\end{figure}

Bulk micromachining is another class of processes for micromechanical electronics systems (MEMS) used in our manufacturing process. For bulk processes, substrate or bulk is removed via etching (either dry or wet) to create microstructures, such as cavities or through-wafer holes, both of which we utilize in our process. Wet etching, where the etchants are liquid, has many benefits over dry etching, where the etchants are plasma. Wet etching is faster, cheaper, more anisotropic, and lower in chemical waste output \cite{BISWAS2006519}. In the case of the slits for KOSMOS, we are using a 35\% concentration of potassium hydroxide (KOH) at 100$^{\circ}$C to etch the silicon. KOH-etching is a well-documented anisotropic Si etch process, which takes advantage of the etching ratio between the crystal planes in the silicon lattice, specifically the $\langle 100 \rangle$ and $\langle 111 \rangle$ planes \cite{SATO199887}. Because KOH preferentially etches the $\langle 100 \rangle$ plane and minimally etches the $\langle 111 \rangle$ plane, the $\langle 111 \rangle$ plane is exposed as the KOH etches through the wafer (Figure \ref{fig:wafer}). This creates a rectangular-shaped cavity with slanted sidewalls, which are the $\langle 111 \rangle$ planes. Because of this crystalline structure of silicon, the surface of each wafer is atomically flat, and through anisotropic etching with KOH, the etched $\langle 111 \rangle$ surfaces have an estimated roughness of 3-5 nm \cite{Ezoe:06}. In Section 4, we observe the maximum roughness of this plane for the slits, with results summarized in Table \ref{tab:ref-res}.

We explored and refined a known method of KOH (Potassium Hydroxide) silicon etching to precisely etch through silicon wafers. The method below was initially developed and tested by engineers at the Washington Nanofabrication Facility (WNF) and Ref. \citenum{Lozo2017}, based on work done by Ref. \citenum{10.1117/12.672369}. For specifics on tools used, their specifications, etc., see the manual \cite{Tran2021}. Below we present a general overview of the approach and results. 

\begin{figure}
\begin{center}
\begin{tabular}{c}
\includegraphics[width=1\textwidth]{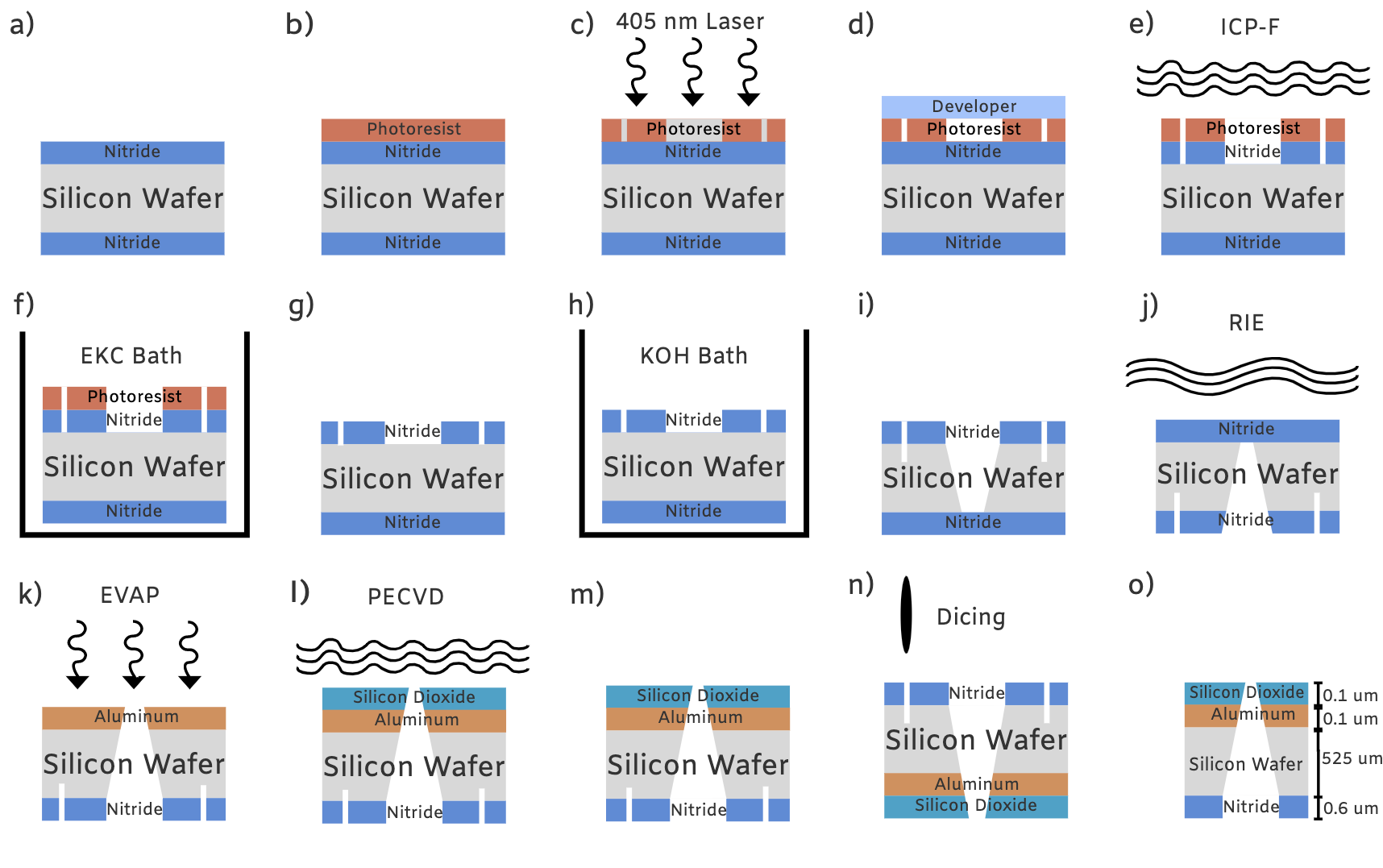} 
\end{tabular}
\end{center}
\caption 
{ \label{fig:fabprocess}
Schematic of fabrication process steps. Thicknesses not to scale. a) Starting silicon wafer, coated on both sides with silicon nitride. b) Coated with photoresist. c) Photoresist written with 405 nm laser. d) Photoresist developed. e) ICP-F etches the exposed silicon nitride. f) EKC bath to remove photoresist. g) Pattern from photoresist transferred onto the nitride. h) KOH bath to etch exposed silicon. i) Slit and dicing lanes etched into silicon wafer. j) Nitride at the bottom of wafer etched off. k) Aluminum evaporated onto wafer. l) Silicon dioxide deposited onto aluminum. m) Light can now go through the slit and have whatever light that doesn't go through reflected back. n) Dicing the extra parts of the wafer off. o) Now completed slit ready for mounting. Silicon in gray; Nitride in blue; Silicon dioxide in teal; Photoresist in dark orange; Aluminum in orange; Developer in light blue. Note - steps j to l were switched in the initial testing of this process such that the reflective side was the wider side. After testing at APO, future slits will be manufactured as described in this diagram.} 
\end{figure} 

\subsection{Photolithography and Pattern Transfer}
The processing begins with 525 $\mu$m thick, 100 mm diameter silicon wafers coated with 25 $\mu$m LPCVD silicon nitride on both sides by Rogue Valley Microdevices\footnote{For more information, please refer to the \href{https://roguevalleymicrodevices.com/}{Rogue Valley Microdevices Website}} (Figure \ref{fig:fabprocess}a). The nitride acts as a mask to protect the silicon from dissolving in the KOH bath, as silicon nitride etching by KOH is negligible ($<$ 1 nm/hour) if etched at all. To etch the nitride such that the silicon beneath it can be exposed, we used photolithography to create a nitride mask on one side of the wafer for etching the silicon. First, we coated the top of the wafer with 1.5 $\mu$m of AZ 1512 photoresist (Figure \ref{fig:fabprocess}b). We then used a 175W 405nm laser with a 20mm write head (\href{https://heidelberg-instruments.com/product/dwl-66-laser-lithography-system/}{Heidelberg DWL 66+ laser lithography tool}) with edge roughness of 0.11 $\mu$m \cite{heidelberg} to expose the photoresist (Figure \ref{fig:fabprocess}c) with the pattern in Figure \ref{fig:slit-mask-dim}. We chose the 20mm write head due to its faster write speed and because the resulting edge roughness is small compared to the width of the slit.

Then we developed the photoresist with AD10, vacuum baked at 100$^{\circ}$C for 60 seconds (Figure \ref{fig:fabprocess}d). This developed photoresist acts as a mask, allowing us to etch the nitride off where the slit will be. We use ICP (inductively coupled plasma) to remove the nitride with a CHF3-O2 plasma at 20$^{\circ}$C (Figure \ref{fig:fabprocess}e). Bare silicon is thus exposed where the slit and dicing lanes will be. We removed the photoresist in an EKC bath (Figure \ref{fig:fabprocess}f).

\subsection{KOH Etching and Plasma Etching}
With a nitride mask protecting the parts of the silicon wafer we do not want removed (Figure \ref{fig:fabprocess}g), we used the KOH to etch the silicon to create the slit (Figure \ref{fig:fabprocess}h). In our process, this took between 5-14 hours (further discussed in Section \ref{ssec:Advantages}). In ideal conditions, a 35\% KOH solution at 100$^{\circ}$C should etch this in 2 hours \cite{Seidel_1990,BISWAS2006519}. The remaining nitride on the back of the wafer is removed via the reactive ion etcher (Figure \ref{fig:fabprocess}j).
 
\subsection{Metallization and Assembly}
We then evaporated (through physical vapor deposition) 1 $\mu$m on aluminum onto the side of the wafer with the narrow end of the slit to make it more reflective (Figure \ref{fig:fabprocess}k). To protect the aluminum from oxidation (which causes the aluminum to become dull and less reflective), we used plasma-enhanced vapor deposition (PECVD) to deposit 100 nm of silicon dioxide onto the wafer (Figure \ref{fig:fabprocess}l). Lastly, we diced the wafers into rectangles for the mounts using a Disco diamond dicing saw (Figure \ref{fig:fabprocess}n; Figure \ref{fig:slit-mask-dim}). Three of each size slit were mounted for use, and the remaining are kept as extras. 

\section{Data and Analysis} 
\label{sect:Data}

To understand whether and/or how the uniformity in width across each slit and the average roughness of the slits impact spectral data, we must first characterize the width and roughness of each slit. However, due to the steep, deep sidewalls of the slits, which are very narrow, it is not possible to use traditional tools to measure the roughness of the slits, such as a profilometer. Instead, we must image the slits and define roughness and width from imaged data.

After measuring the characteristics of each slit, we used KOSMOS in lab to measure the potential impacts, scattering in particular, of the roughness and uniformity on spectral data in a controlled setting. Finally, in our future work, we will use KOSMOS on sky to understand the performance of the slits. We will characterize the throughput and the spectrophotometric data quality as a function of position on the slit, in conditions that observers will typically experience.

\subsection{Acquisition - Microscopic Data}
\begin{figure}
\begin{center}
\begin{tabular}{c}
\includegraphics[scale=.3]{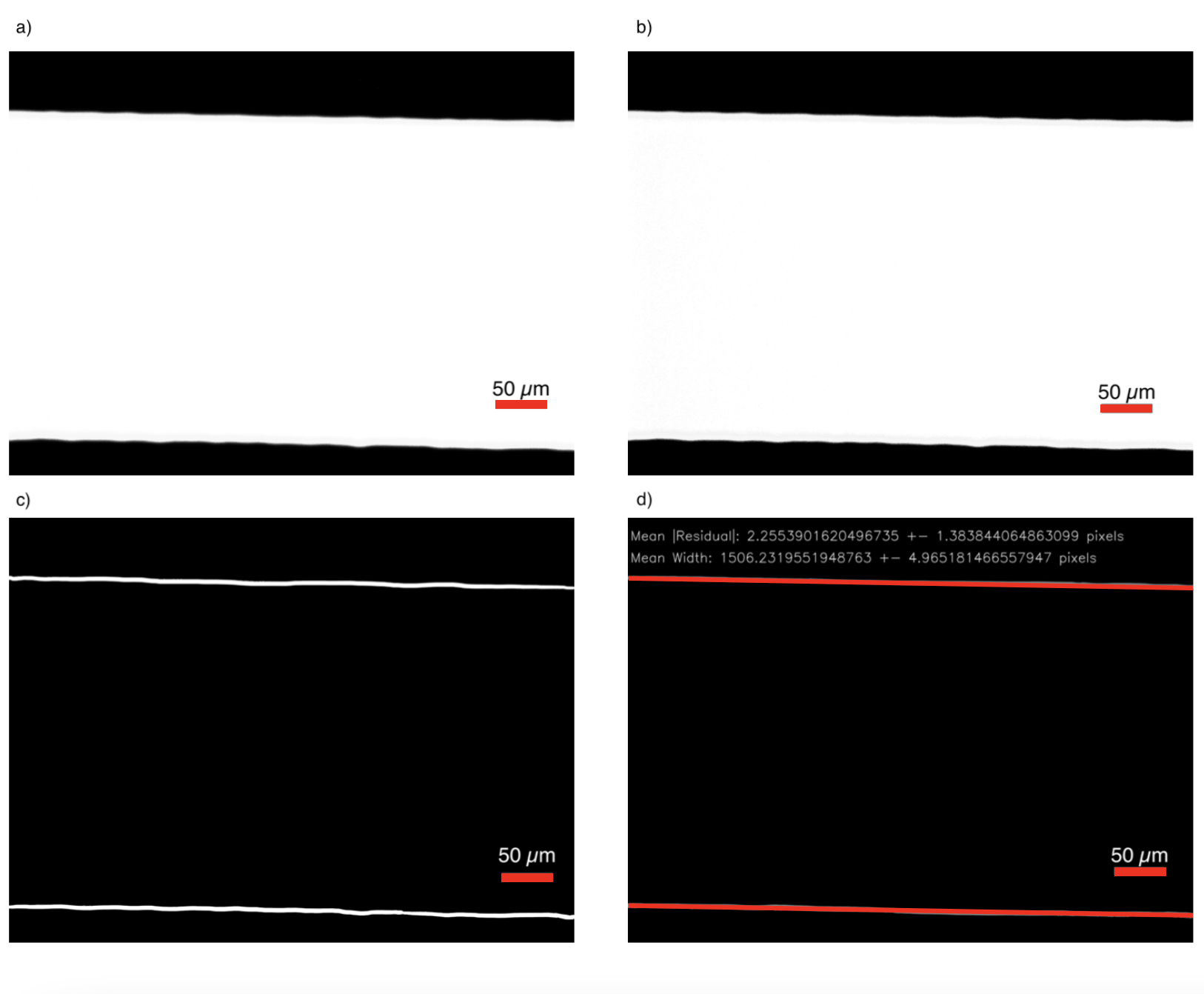} 
\end{tabular}
\end{center}
\caption{
\label{fig:slit-proc}
Sample of data taken on the microscope at various stages of metrology, specifically a 5" reflective slit at 20x magnification with light coming through the reflective side of the slit and hitting the eye piece and detector (backlit). a) Data taken with microscope in uncompressed .tiff file format. The yellow is a result of the scaling of the intensity of the image. b) Data after unsharp masking. c) Edges detected by Canny edge detection in OpenCV. d) Straight lines (in red) detected in data using probabilistic Hough transform with residuals calculated by taking the separation between the detected edge (in white) and the detected line (in red). The text on the image is the width and roughness output for just this image. Note: The detected edges and lines are exaggerated in c) and d) to improve readability. The 50 micron scale bar was added retroactively for sense of scale.}
\end{figure}

For each slit, ten sections along the slit were imaged randomly (Figure \ref{fig:slit-proc}a), taken on an optical Leica DM compound microscope, with the slit being illuminated from below with the light pointing through the slit towards the camera. For chemically-etched slits with slit widths between 60 $\mu$m and 400 $\mu$m, images were taken at magnifications of 5x, 10x, and 20x. Images were taken with the light illuminating through the narrow side of the slits and the wide side of the slits to test whether there is a measurable difference in the widths of the slit. However, as the narrow side of the slit, which is its targeted width, sets the amount of light coming into the slit, the measured slit widths ended up being the same on both sides.  
For etched slits with slit widths larger than 1 mm, images were taken at 5x, as the slit is too wide for imaging at higher magnifications. For wire EDM slits, data were taken with the 10x objective to allow for standardized comparisons between all slits.

Using numerical apertures and and estimated peak wavelength to be at 550 nm, we obtain the resolution of images at each objective. We also imaged a reticle at 5x, 10x, and 20x magnifications to find the conversion of pixels to microns Table(\ref{tab:micro-objective}). Due to the low spatial resolution of the optical microscope and unresolved structure, these measurements set upper limits on the roughness of the slits and high errors on the measured width.

\begin{table}[h!]
\renewcommand{\thetable}{\arabic{table}}
\centering
\caption{Spatial Resolution of Each Microscope Objective} \label{tab:micro-objective}
\begin{tabular}{cccc}
\hline
\hline
Objective & Numerical Aperture & Resolution [$\mu$m] & Pixel Scale [$\mu$m/pix]\\
\hline
5x & 0.12 & 2.29 & 0.9 $\pm$ 0.1\\
10x & 0.30 & 0.92 & 0.45 $\pm$ 0.02\\
20x & 0.40 & 0.69 & 0.22 $\pm$ 0.01\\
\hline
\end{tabular}
\end{table}

\subsection{Data Analysis - Microscopic Data}\label{ssec:OpenCV}
\begin{figure}
\begin{center}
\begin{tabular}{c}
\includegraphics[scale=.2]{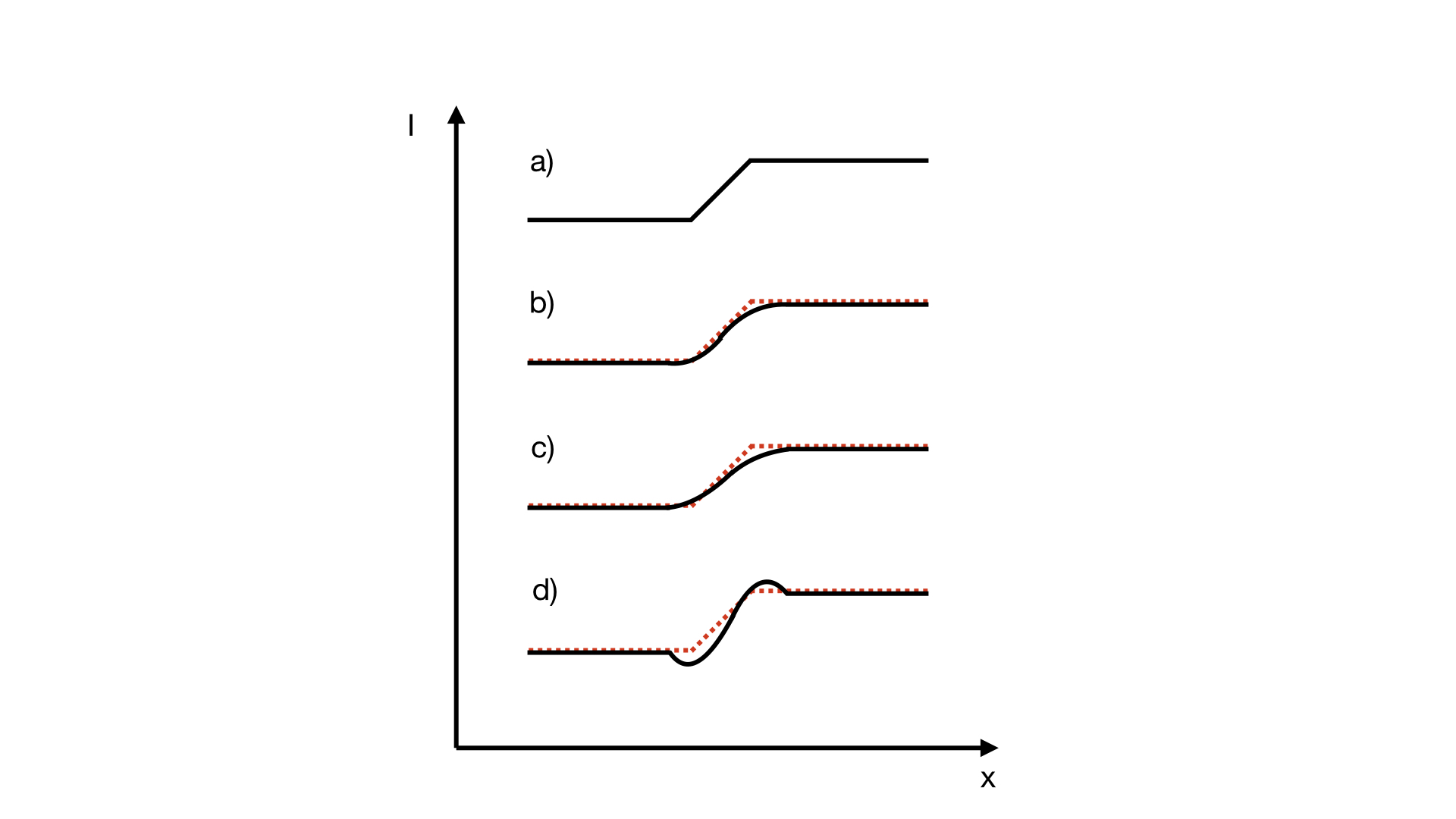} 
\end{tabular}
\end{center}
\caption 
{ \label{fig:unsharp}
Cross-sections of edges in the process of unsharp masking a) idealized edge; b) observed edge; c) smoothed version of b)-  see Eq. \ref{eq:f_smooth}; d) sharpened edge as a result of subtracting a proportion of c) from b)- see Eq. \ref{eq:f_sharpened}} 
\end{figure}

Images were sharpened using Gaussian unsharp masking (Figure \ref{fig:slit-proc}b), a common digital signal processing technique used to enhance the intensity at the edges \cite{rosenfeld_1969}. First, we smooth (Equation \ref{eq:f_smooth}) the original image by convolving a Gaussian kernel (Equation \ref{eq:Gaussian_kernel}), which is a discretized approximation of a 2-D Gaussian- in this case in a 5x5 matrix which acts as a highpass filter, with the original image. Because the smoothing is done at a smaller spatial scale than the rough features (more than 5x5 pixels), resolution of the slit edges are not impacted.
\begin{equation}
    f_{smoothed}(x,y) = G(x,y) * f(x,y)  
    \label{eq:f_smooth}
\end{equation}
$f_{smoothed}(x,y)$ (Figure \ref{fig:unsharp}c) is the smoothed or blurred version of the original input image $f(x,y)$ (Figure \ref{fig:unsharp}b). $G(x,y)$ is the discrete approximation for a 2-D Gaussian function, where x and y are pixel position and $\sigma$ is the standard deviaiton of the Gaussian function:
\begin{equation}
    G(x,y) = \frac{1}{\sqrt{2\pi\sigma^2}}e^{-\frac{x^2+y^2}{2\sigma^2}} 
    \label{eq:Gaussian_kernel}
\end{equation}
Then one subtracts a blurred or smoothed image (Equation \ref{eq:f_smooth}) from the original image resulting in a sharpened image with reduced Gaussian noise (Figure \ref{fig:unsharp}d; Figure \ref{fig:slit-proc}b; Equation \ref{eq:f_sharpened}) \cite{schalkoff_1989}. 
\begin{equation}
    f_{sharpened}(x,y) = f(x,y) - f_{smoothed}(x,y)
    \label{eq:f_sharpened}
\end{equation}
This is a vital step for allowing Canny, the edge detection algorithm used and described below, to properly detect the edges of the slit.

\begin{figure}
\begin{center}
\begin{tabular}{c}
\includegraphics[scale=.4]{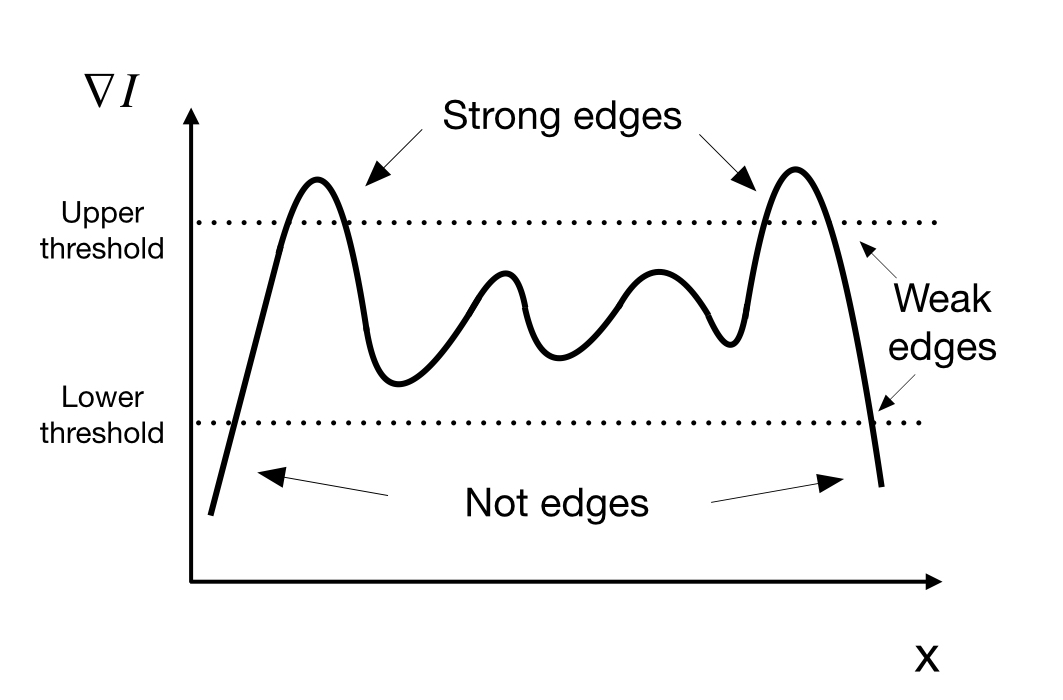} 
\end{tabular}
\end{center}
\caption 
{ \label{fig:canny}
One-dimensional example of how Canny detects strong edges above the upper threshold and connects the weak edges between the strong edges to make one detected edge. Any pixels with intensity gradient below the lower threshold is not detected as an edge.}
\end{figure}

Images were then analyzed with algorithms from \href{https://opencv.org/about/}{OpenCV}, an open-source computer vision and machine learning software library. The edges of the slit were detected with OpenCV's implementation of the Canny edge detector, a widely-used edge detection algorithm \cite{canny} (Figure \ref{fig:slit-proc}c). After Gaussian unsharp masking, the algorithm finds the intensity gradient at each pixel in the image by filtering with a Sobel kernel, outputting the first derivative of intensity in both horizontal and vertical directions. With the vertical and horizontal first derivatives, it computes magnitude and direction of the gradient at each pixel location. It then searches for potential edges by looking for high gradients in brightness. Then, it applies an upper and lower threshold, where it rejects all potential edges below the lower gradient threshold, accepts all strong edges above the upper threshold, and tentatively accepts weak edges in between the upper and lower thresholds. Lower thresholds were determined by testing various input values to check whether Canny would successfully detect the edges of the slit without detecting any background Poisson noise or non-existent structure as edges. Finally, using hysteresis, any strong edge with weak edges in between are connected together to be one edge, and all weak edges not connected to strong edges are suppressed, shown in Figure \ref{fig:canny}. With a picture of what the edges of the slit looks like from Canny (Figure \ref{fig:slit-proc}c), we then use Hough transforms to fit a line to the edge.

\begin{figure}[h!]
\begin{center}
\begin{tabular}{c}
\includegraphics[scale=.25]{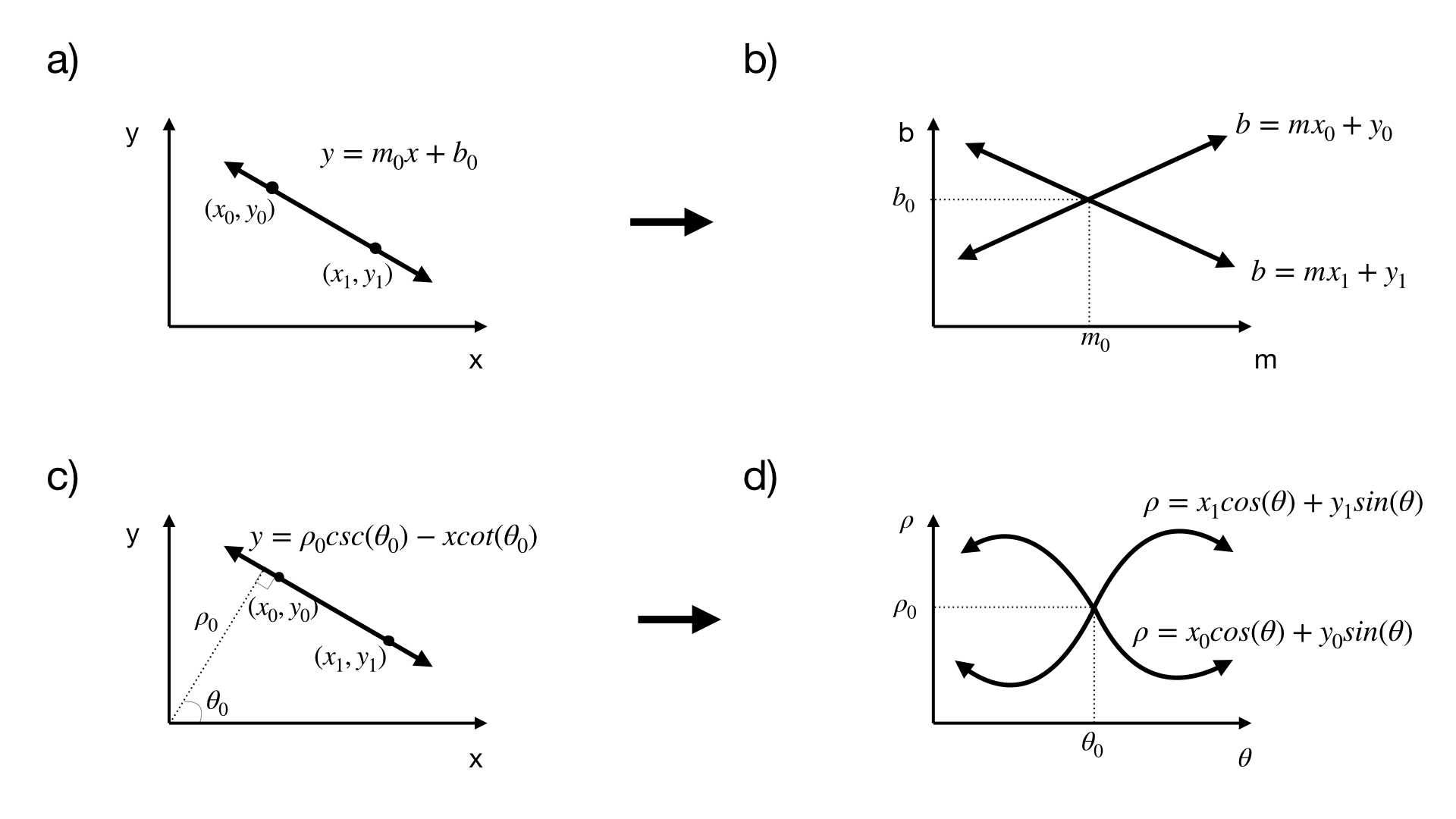} 
\end{tabular}
\end{center}
\caption 
{\label{fig:houghspace}
a) Representation of a line in xy-space where $m_0$ is the slope and $b_0$ is the y-intercept. b) Representation of a line in mb-space, which is a point where lines $(x_0,y_0)$ and $(x_1,y_1)$ intersect. c) Representation of a line xy-space where $\rho_0$ is the distance between the origin and the line and $\theta_0$ is the angle between the orthogonal to the line and x-axis. d) Representation of a line in Hough space - looks like a point where waves $(x_0,y_0)$ and $(x_1,y_1)$ intersect.}
\end{figure}

Straight lines were fit to the detected edges using OpenCV's probabilistic Hough transform algorithm (HoughLinesP) (Figure \ref{fig:slit-proc}d). The probabilistic or randomized Hough transform \cite{XU1990331} is a modification of the Hough transform \cite{BALLARD1981111,10.1145/361237.361242,houghpatent}. This allows us to fit straight lines to those detected edges to understand how far the deviation of the slit is from an idealized straight edge. HoughLinesP randomly picks some number of points and maps it in the Hough parameter space, $\theta$ and $\rho$, where $\theta$ is an alternative way to measure the slope of a line and $\rho$ measures the distance of the line to the origin (Figure \ref{fig:houghspace}). HoughLinesP then searches a grid of $\rho$ and $\theta$ to find the best-fit line, given by the line with $\rho$ and $\theta$ that has the most votes or points that fall in that line. 

The roughness is determined by the mean residual (absolute value of the difference) between the edge detected and the best-fit line determined by HoughTransformP along the sampled image. The slit width is the average distance between the two lines detected. Due to the reflective slits being angled at 15$^{\circ}$ for the slitviewer, the effective slit width is slightly smaller (by cos(15$^{\circ}$)) than the measured width. The width uniformity is the standard deviation of the measured width along the length of the slit. These roughness and width measurements for both matte and reflective slits, as well as the effective width for the reflective slits, are presented in the Section \ref{sect:Discussion}. 

\subsection{Acquisition - KOSMOS Data - In Lab} 
For each mounted slit, both chemically-etched and wire EDM, spectra were taken using the krypton, neon, and argon internal calibration lamps on simultaneously, alternating between red and blue grisms. By having all internal calibration lamps on, more emission lines were present across the entire optical bandpass and minimized the need to wait for warm up times of the individual calibration lamps for each set of observations. Both red and blue grisms were separately used in observations to gauge impacts on throughput throughout the entire bandpass. Darks, or images taken without opening the shutter, were taken to characterize the thermal noise of the detector. Flats, which characterize the individual pixel response of the KOSMOS CCD (charge coupled device) detector, were not taken. The spectra for wire EDM and chemically-etched slits of the same size and position fell in roughly the same place on the detector, which was sufficient for sample data to compare between the machined and chemically-etched slits. For specific details on KOSMOS as an instrument, see Kadlec et al., in prep.

\subsection{Data Analysis- KOSMOS Data - In Lab} 
Data were reduced by subtracting a median-combined master dark, where all the darks taken were median-combined, from spectral data. From this reduced data, we divided by exposure time to convert the measured counts in analog-digital units (ADU) to flux, in units of ADU/s/pixel. The background, both background from outside of the slit and spectral background, was then averaged in each frame. The background light through the slits provides an analog for scattering, though not absolute, is sufficient for comparisons of scattering in the matte and reflective slits. Since data were taken with both the red and blue dispersers for each mounted slit, the background and slit widths obtained from the separate red and blue disperser frames were averaged.
\begin{figure}
    \centering
    \includegraphics[width=\textwidth]{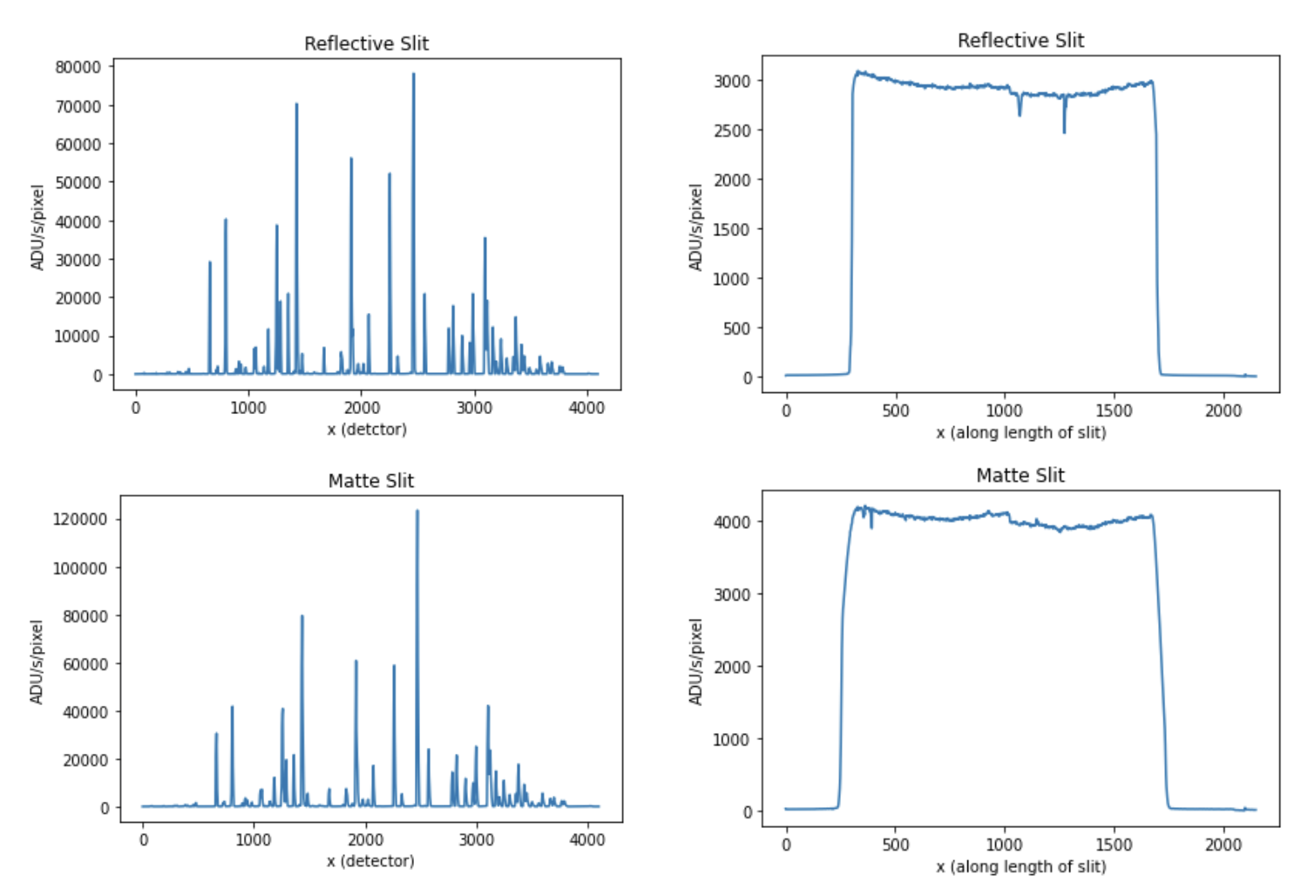}
    \caption{Top left: Example spectra taken using the chemically etched, reflective slit; Top right: Example slit cross section, reflective slit; Bottom left: Example spectra taken using the wire EDM, matte slit; Bottom right: Example slit cross section, matte slit}
    \label{fig:example-spectra}
\end{figure}

\section{Results and Discussion} \label{sect:Discussion}
Due to the wide variety of scientific drivers for instruments, which are infrequently built, it is difficult to disentangle oral history from holdovers from older technologies. Below we discuss the results of a thorough examination of slit roughness and uniformity, as we want to be as cautious as possible about reducing possible sources of background and increasing throughput in science data. As new technologies become available, we have taken the opportunity to measure and compare to older techniques. By characterizing the machined and chemically-etched slits, we can fully understand the performance limits of these slits. This way, we have all the information to judge the trade-offs and what can be improved and what we should continue to use.

\subsection{Slit Results- Width Uniformity and Maximum Roughness}
By comparing the results in Tables \ref{tab:ref-res} and \ref{tab:matte-res}, the reflective wet-etched slits are less rough, with smaller residuals than the matte machined slits, throughout the various target sizes of slits. The overall average roughness of chemically-etched slits across all slit widths is 0.42 $\pm$ 0.03 $\mu$m. The average variation across etched slits (average of the standard deviation of each slit), excluding the widest slits, which could only be taken at 5x, setting errors at 20\%) is 6 $\pm$ 3 $\mu$m. The overall average roughness of machined slits is 1.06 $\pm$ 0.04 $\mu$m. The average variation across machined slits is 10 $\pm$ 6 $\mu$m. This makes the etched slits on average 2.5 times smoother than the machined slits on average. Due to the low resolution of the optical microscope used for measurements, and subsequent error propagation in the measured width, the comparison of width uniformity between the two types of slits is inconclusive.

\begin{figure}[h!]
    \centering
    \includegraphics{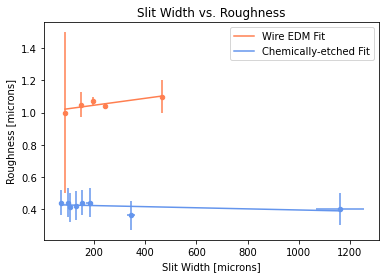}
    \caption{Plot of average roughness as a function of average width with error bars from the standard deviation of both. The roughness of the machined slits is consistently higher than that of the etched slits. Both have no width dependence.}
    \label{fig:roughness_v_width}
\end{figure}

To examine the roughness as a function of width for both methods of fabrication, we fit roughness versus width in Figure \ref{fig:roughness_v_width}. The roughness of the slits do not depend on widths in both the machined and etched slits. Additionally, the wire EDM slits are consistently rougher than chemically-etched slits, no matter the slit width.

\begin{table}[h!]
\renewcommand{\thetable}{\arabic{table}}
\centering
\caption{Average Reflective Slits Roughness/Residuals} \label{tab:ref-res}
\begin{tabular}{ccc}
\hline
\hline
Width ["] & Width [$\mu$m] & Average Roughness/Residuals [$\mu$m]\\
\hline
0.5 & 29.2 & 0.44 $\pm$ 0.08 \\
0.8 & 46.8 & 0.44 $\pm$ 0.09 \\
1.0 & 58.5 & 0.41 $\pm$ 0.09 \\
1.3 & 76.0 & 0.42 $\pm$ 0.09 \\
1.6 & 93.6 & 0.44 $\pm$ 0.08 \\
2.0 & 117.0 & 0.44 $\pm$ 0.09\\
5.0 & 292.4 & 0.36 $\pm$ 0.09 \\
20.0 & 1169.6 & 0.4 $\pm$ 0.1 \\
\hline
\end{tabular}
\end{table}

\begin{table}[h!]
\renewcommand{\thetable}{\arabic{table}}
\centering
\caption{Average Matte Slits Roughness/Residuals} \label{tab:matte-res}
\begin{tabular}{ccc}
\hline
\hline
Width ["] & Width [$\mu$m]& Average Roughness/Residuals [$\mu$m]\\
\hline
0.584 & 91 & 1.0 $\pm$ 0.5  \\
0.876 & 137 & 1.05 $\pm$ 0.08 \\
1.168 & 183 & 1.07 $\pm$ 0.03 \\
1.46 & 228 & 1.04 $\pm$ 0.02 \\
2.92 & 456 & 1.1 $\pm$ 0.1  \\
\hline
\end{tabular}
\end{table}

\begin{table}[h!]
\renewcommand{\thetable}{\arabic{table}}
\centering
\caption{Individual Reflective Slit Widths} 
\label{tab:ref-width}
\begin{tabular}{cccc}
\hline
\hline
Measured Width [$\mu$m] & Effective Width ["]\\
\hline
74 $\pm$ 5  & 0.42 $\pm$ 0.03\\
64 $\pm$ 5 & 0.36 $\pm$ 0.03\\
70 $\pm$ 3 & 0.40 $\pm$ 0.02  \\
75 $\pm$ 3 & 0.42 $\pm$ 0.02  \\
75 $\pm$ 2 & 0.42 $\pm$ 0.01 \\
81 $\pm$ 6 & 0.46 $\pm$ 0.04 \\
\hline
95 $\pm$ 4 & 0.54 $\pm$ 0.03 \\
99 $\pm$ 4 & 0.56 $\pm$ 0.03 \\
102 $\pm$ 7 & 0.58 $\pm$ 0.04 \\
102 $\pm$ 5 & 0.58 $\pm$ 0.03\\
104 $\pm$ 4 & 0.59 $\pm$ 0.03 \\
\hline
107 $\pm$ 5 & 0.60 $\pm$ 0.03 \\
116 $\pm$ 7 & 0.65 $\pm$ 0.04 \\\
116 $\pm$ 8 & 0.66 $\pm$ 0.05 \\
\hline
130 $\pm$ 6 & 0.73 $\pm$ 0.03 \\ 
133 $\pm$ 5 & 0.75 $\pm$ 0.03\\
133 $\pm$ 7 & 0.75 $\pm$ 0.04 \\
130 $\pm$ 10 & 0.73 $\pm$ 0.07\\
141 $\pm$ 4 & 0.79 $\pm$ 0.04 \\
136 $\pm$ 4 & 0.77 $\pm$ 0.02  \\
\hline
150 $\pm$ 10 & 0.86 $\pm$ 0.06\\
154 $\pm$ 6 & 0.87 $\pm$ 0.03\\
155 $\pm$ 6 & 0.88 $\pm$ 0.04\\
147 $\pm$ 6 & 0.83 $\pm$ 0.04\\
162 $\pm$ 4 & 0.92 $\pm$ 0.03 \\
\hline
178 $\pm$ 4 & 1.01 $\pm$ 0.02 \\
178 $\pm$ 5 & 1.01 $\pm$ 0.05 \\
210 $\pm$ 10 & 1.18 $\pm$ 0.07 \\
\hline
360 $\pm$ 10 & 2.05 $\pm$ 0.09 \\
370 $\pm$ 10 & 2.10 $\pm$ 0.06\\
350 $\pm$ 20 & 2.0 $\pm$ 0.1 \\
\hline
1250 $\pm$ 140 & 7.3 $\pm$ 0.8 \\
1020 $\pm$ 120 & 6.0 $\pm$ 0.7 \\
1139 $\pm$ 130 & 6.7 $\pm$ 0.8 \\
1240 $\pm$ 140 & 7.3 $\pm$ 0.8\\
\hline\\
\end{tabular}\\
\small
NOTE: As stated in Section \ref{sect:Design}, target widths were calculated with the incorrect plate scale. Instead, we are grouping the fabricated slits in sets with similar slit widths as demarcated by the horizontal lines above, showing measurements made in microns and converted to arcseconds using the correct plate scale. 
\end{table}

\begin{table}[h!]
\renewcommand{\thetable}{\arabic{table}}
\centering
\caption{Individual Matte Slit Width} \label{tab:matte-width}
\begin{tabular}{cc}
\hline
\hline
Measured Width [$\mu$m] & Measured Width ["]\\
\hline
94 $\pm$ 4 & 0.55 $\pm$ 0.02\\
96 $\pm$ 4 & 0.56 $\pm$ 0.02\\
69 $\pm$ 3 & 0.40 $\pm$ 0.02\\
\hline
137 $\pm$ 6 & 0.88 $\pm$ 0.04\\
148 $\pm$ 7 & 0.87 $\pm$ 0.04\\
158 $\pm$ 7 & 0.92 $\pm$ 0.04\\
\hline
196 $\pm$ 9 & 1.15 $\pm$ 0.05\\
200 $\pm$ 9 & 1.17 $\pm$ 0.05\\
192 $\pm$ 9 & 1.12 $\pm$ 0.05\\
\hline
247 $\pm$ 11 & 1.45 $\pm$ 0.06\\
241 $\pm$ 11 & 1.41 $\pm$ 0.06\\
245 $\pm$ 11 & 1.43 $\pm$ 0.06\\
\hline
468 $\pm$ 21 & 2.7 $\pm$ 0.1\\
468 $\pm$ 21 & 2.7 $\pm$ 0.1\\
462 $\pm$ 21 & 2.7 $\pm$ 0.1\\
\hline
\end{tabular}
\end{table}

\begin{figure}
    \centering
    \includegraphics[width=\textwidth]{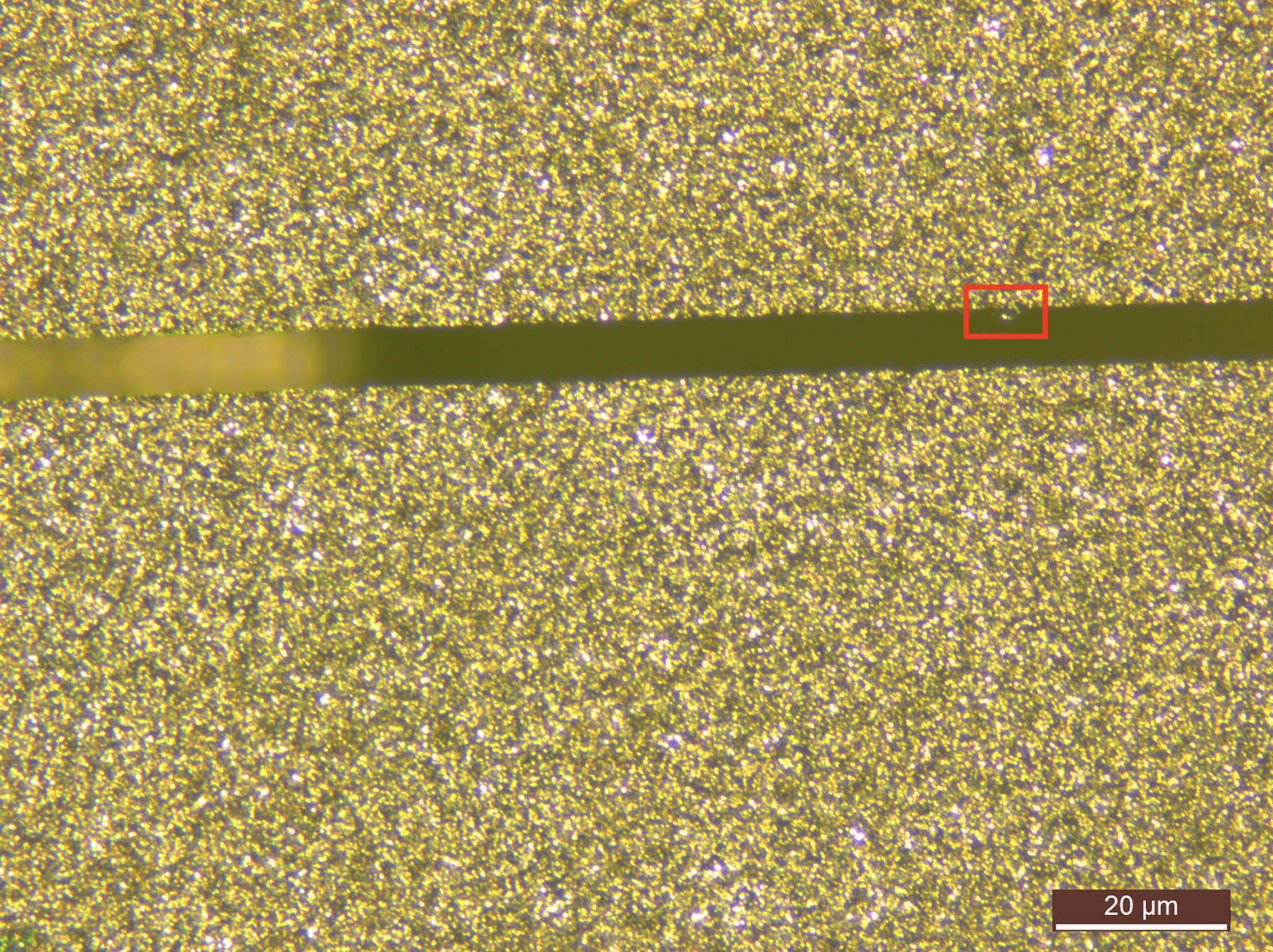}
    \caption{Imaged slit, where the nitride is the glittery gold and the slit is in the middle. The red box highlights a bit of remaining nitride on the slit. For scale, a 20 micron-scale bar was added.}
    \label{fig:remaining-nitride}
\end{figure}

In comparing Tables \ref{tab:ref-width} and \ref{tab:matte-width}, the chemically-etched slits also are more uniform in width throughout the individual slits, as the error is more consistent and generally smaller than machined slit width error. This is due to the 0.11 $\mu$m precision of the laser lithography tool, which sets the shape of the slit. Under ideal conditions, the KOH etching sets the size of the slits and will etch a smooth edge in the silicon. However, due to the remaining nitride on the slit (Figure \ref{fig:remaining-nitride}) as well as residue from the dicing tape being removed with acetone seen when visually inspected, the slits are rougher than expected. The reactive ion etching (RIE) etcher responsible for removing the remainder of the nitride on the slit did not appear to remove all of it.

In the testing of this process, we found a large disparity between the targeted slit width and the measured slit width of the chemically-etched slits. This comes from the unpredictability of our wet-etching process, which was missing a step of removing silicon dioxide prior to the KOH etch, discussed in Section \ref{ssec:Advantages}. Instead of grouping them by targeted slit width, we chose to demarcate the slits by their functional measured slit width for practical purposes in Table \ref{tab:ref-width}.

\subsection{Slit Results - Scattering}

\begin{figure}
    \centering
    \includegraphics{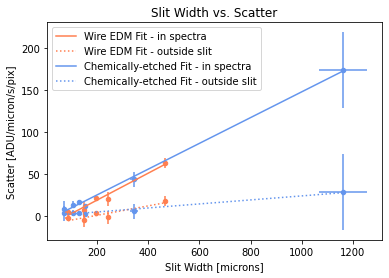}
    \caption{Plot of average scattering vs average width with standard deviations of both as error bars. Scattering was measured with two methods, with the solid line fit representing scatter measured in the spectra and the dotted line fit representing scattered light outside the slit. Scattering increases linearly as a function of width for both wire EDM and chemically-etched slits with both methods.}
    \label{fig:scatter_v_width}
\end{figure}

In examining Tables \ref{tab:ref-scatter} and \ref{tab:matte-scatter}, the scattering of the chemically-etched slits is higher on average than that of the machined slits. Since the etched slits are smoother than the machined slits, the reflectivity, not the roughness, is the major contributor to the observed scattering. The scattering increases as slit width increases for both matte and reflective slits and is greater for reflective slits (Figure \ref{fig:scatter_v_width}). To normalize the scatter by slit width and compare the impact of reflectivity, the scatter was divided by the width, giving units of ADU/pix/s/micron. The scatter/width for matte slits including continuum is 0.09 $\pm$ 0.5 ADU/pix/s/micron and 0.13 $\pm$ 0.3 ADU/pix/s/micron for reflective slits. For light outside the slit, the scatter/width for matte slits is -0.0004 $\pm$ 0.03 ADU/pix/s/micron and 0.04 $\pm$ 0.02 ADU/pix/s/micron for reflective slits. In both scenarios, the scatter using the reflective slits is higher, but with proper background subtraction, the impacts of the scatter can be minimized.

\begin{table}[h!]
\renewcommand{\thetable}{\arabic{table}}
\centering
\caption{Average Reflective Slit Scatter} \label{tab:ref-scatter}
\begin{tabular}{ccc}
\hline
\hline
Measured Width & Scatter in Spectra & Scatter Outside Slit \\
{[$\mu$m]} &  [ADU/pix/s] &  [ADU/pix/s]\\
\hline
74 $\pm$ 5 & 11.4 $\pm$ 0.9 & 4 $\pm$ 1\\
64 $\pm$ 5 & 11 $\pm$ 5 & 6 $\pm$ 4\\
75 $\pm$ 2 & 6 $\pm$ 1 & 3 $\pm$ 1 \\
75 $\pm$ 3 & 7 $\pm$ 2 & 4 $\pm$ 1  \\
70 $\pm$ 3 & 8 $\pm$ 3 & 3 $\pm$ 1  \\
\hline
95 $\pm$ 4 & 14 $\pm$ 3 & 4 $\pm$ 1 \\
99 $\pm$ 4 & 21 $\pm$ 5 & 10 $\pm$ 6 \\
104 $\pm$ 4 & 16 $\pm$ 1 & 5 $\pm$ 2 \\
116 $\pm$ 8 & 10 $\pm$ 2 & 4 $\pm$ 1 \\
102 $\pm$ 7 & 13 $\pm$ 5 & 4 $\pm$ 1 \\
102 $\pm$ 5 & 11 $\pm$ 4 & 2 $\pm$ 1\\
\hline
133 $\pm$ 7 & 14 $\pm$ 6 & 2.6 $\pm$ 0.9 \\
130 $\pm$ 10 & 18.7 $\pm$ 0.6 & 4 $\pm$ 1\\
130 $\pm$ 6 & 18.2 $\pm$ 0.1 & 4 $\pm$ 2 \\ 
133 $\pm$ 5 & 19.0 $\pm$ 0.2 & 5 $\pm$ 2\\
\hline
150 $\pm$ 10 & 10 $\pm$ 2 & 2.9 $\pm$ 0.8\\
154 $\pm$ 6 & 19 $\pm$ 2 & 4 $\pm$ 0.7\\
147 $\pm$ 6 & 15 $\pm$ 5 & 3 $\pm$ 2\\
\hline
360 $\pm$ 10 & 36 $\pm$ 11 & 4 $\pm$ 1 \\
370 $\pm$ 10 & 39 $\pm$ 11 & 8.6 $\pm$ 0.3\\
350 $\pm$ 20 & 56 $\pm$ 4 & 5 $\pm$ 3 \\
\hline
1250 $\pm$ 140 &  188 $\pm$ 14 & 35 $\pm$ 3 \\
1139 $\pm$ 130 & 113 $\pm$ 33 & 28 $\pm$ 1 \\
1240 $\pm$ 140 & 221 $\pm$ 40 & 25 $\pm$ 7 \\
\hline
\end{tabular}\\
\small NOTE: Not all slits were mounted as some were made as extras, so the table of data taken with final library of slits is shorter than the other table. 
\end{table}

\begin{table}[h!]
\renewcommand{\thetable}{\arabic{table}}
\centering
\caption{Average Matte Slits Scattering} \label{tab:matte-scatter}
\begin{tabular}{ccccc}
\hline
\hline
Target Width & Target Width & Measured Width & Scatter in Spectra & Scatter Outside Slit \\
{["]} & [$\mu$m] & [$\mu$m] &  [ADU/pix/s] &  [ADU/pix/s]\\
\hline
0.584 & 91 & 94 $\pm$ 4 & 7 $\pm$ 2 & -2 $\pm$ 1\\ 
0.584 & 91 & 96 $\pm$ 4 &7 $\pm$ 2 & -2.0 $\pm$ 0.7\\ 
0.584 & 91 & 69 $\pm$ 3 & 0.2 $\pm$ 1 & -3 $\pm$ 1\\ 
\hline
0.876 & 137 & 137 $\pm$ 6 & -4 $\pm$ 7 & -9 $\pm$ 7\\ 
0.876 & 137 & 148 $\pm$ 7 & 14 $\pm$ 2 & -1 $\pm$ 1\\ 
0.876 & 137 & 158 $\pm$ 7 & 15 $\pm$ 3 & -0.6 $\pm$ 0.2\\ 
\hline
1.168 & 183 & 196 $\pm$ 9 & 21 $\pm$ 10 & 10 $\pm$ 6\\ 
1.168 & 183 & 200 $\pm$ 9 & 25 $\pm$ 1 & 2.4 $\pm$ 0.7\\ 
1.168 & 183 & 192 $\pm$ 9 & 19 $\pm$ 3 & -1.0 $\pm$ 0.5\\ 
\hline
1.46 & 228 & 247 $\pm$ 11 & 10 $\pm$ 1 & -1 $\pm$ 3\\ 
1.46 & 228 & 241 $\pm$ 11 & 27 $\pm$ 3 & -2 $\pm$ 2\\ 
1.46 & 228 & 245 $\pm$ 11 & 25 $\pm$ 5 & -0.7 $\pm$ 0.9\\ 
\hline
2.92 & 456 & 468 $\pm$ 21 & 59 $\pm$ 28 & 30 $\pm$ 19\\ 
2.92 & 456 & 468 $\pm$ 21 & 72 $\pm$ 3 & 15 $\pm$ 2\\
2.92 & 456 & 462 $\pm$ 21 & 61 $\pm$ 3 & 8 $\pm$ 2\\ 
\hline
\end{tabular}
\end{table}

\subsection{Potential Biases and Shortcomings of Measurement Technique}
The measurements for roughness and width were made using an optical microscope with low resolution. Due to this, the measurements made in this paper are upper limits and only meant to be used as a comparison between the two types of slits presented, taken with the same optical microscope. As discussed in Section \ref{sec:conclusions}, measurements of future slits will be made with a scanning electron microscope to better characterize the roughness.

\subsection{Advantages and Disadvantages of Wet Chemical Etching for Slit Fabrication} \label{ssec:Advantages}
The primary advantages of using wet chemical etching to make slits are that they are less rough than that of wire EDM slits. There are many academic labs with nanofabrication tools available, making this fabrication method fairly available. The process for setting the widths and features of the slits is also easy, as it just requires editing a CAD file. In addition, as mentioned in Ref. \citenum{10.1117/12.672369}, these slits are ultra-flat, which makes them more ideal mirrors for the purposes of the slitviewer and minimizes any distortions due to warping or uneven polishing. 

The disadvantage for this specific project was the unpredictability in the etching process. The etch rate was inconsistent across wafers in the bath due to native oxide on the silicon slowing down the KOH etch. A way to mitigate these inconsistencies in future work would be to use BOE (buffered oxide etchant) to etch away the silicon oxide that grew after etching the nitride off, as done in Ref. \citenum{Rao2017}. This is expected to constrain the etch times to be closer to the theoretical through etch time of 2 hours at 35\% KOH at 100$^{\circ}$C \cite{Seidel_1990}, as opposed to the 5-14 hours that it took for this project. Another factor that contributed to the difficulties of this process were that labels with the different slit sizes in the etch process would etch away, making distinguishing different slits difficult. A suggested alternative method of labeling would be making multiple marks in the etching process that do not blend together. Lastly, the dicing lanes would not always fully etch through, though ultimately with the way the dicing saw was set up, it did not matter too much. 

Overall, this is an effective method for those who need reflective, highly uniform, and flat slits with multiple custom sizes provided that there is the addition of the step of a BOE etch to remove the silicon dioxide prior to the KOH etch to reduce unpredictability in sizing. The lead time is longer, as the uniqueness of each project in these academic labs make the process look vastly different for each user and there are a large number of tools to learn to use. It is possible for this process to be outsourced to contractors at facilities with equipment to handle different sized wafers, though this has not been considered at this point due to the process still being tested. Fortunately many of these steps can be done in batches, cutting some of the time it takes for processing. Most of the uncertainty in etch time can be minimized by doing a quick BOE dip prior to the KOH etch, as well as considering the amount of silicon that will be exposed and etched at a time.

\section{Conclusions and Future Work} \label{sec:conclusions}
Chemically-etched slits are overall smoother than the machined slits. The chemically-etched slits and wire EDM slits produce roughly the same scattering within their measurement errors (see Section 5.2).  The main advantage of the chemically-etched slits is how customizable and easy for each user to specify, as it doesn't depend on wire-gauge. The edges, particularly around the ends of the slits, are much smoother than those of the wire-EDM slits, as that method makes a rough hole where the wire enters. For short slits, users won't have to worry about poor data quality at the ends of the slits.

Roughness is not a primary factor in scattering. Rather reflectivity impacts scattering far more. Background in data increases slightly using the reflective chemically-etched slits. Users can choose whether to use matte slits, as the old KOSMOS slits are still available for selection, or reflective slits depending on their observational needs, i.e. whether they need a slitviewer, whether high signal to noise is desired, and whether the slit width needed is available. 

In the future, we will be characterizing the throughput with KOSMOS on-sky. By testing on sky settings, we can determine how the slits perform in the actual conditions observers will have to deal with. Since we have determined the machined and etched slits' roughness and width with microscope data and the scattering from the laboratory data, the on sky KOSMOS data will allow us to determine how the edge roughness and reflectance will impact throughput. Based on the preliminary results with laboratory data taken with KOSMOS, our hypothesis is that the higher scattering due to the reflectivity of the slits will decrease the signal to noise ratio of spectral data. Compared to the large impact of the reflectivity, the slit edge roughness will have a less noticeable effect, similar to the findings in the laboratory measurements. Below are the tests we will be performing. 

We will be taking a variety of data from well-characterized stars and measuring the throughput from both machined and etched slits. Following the method outlined in Ref. \citenum{Lopez2020}, we will characterize throughput of each slit, by deriving an expected photon rate from the star and comparing that to the measured photon rate on the detector. While these numbers are dependent on the observing conditions, they are sufficient for rough comparisons to the two sets of slits. This methodology allows us to more precisely understand the conditions that the data are taken in, as opposed to those in the manufacturer-provided specification sheets for wavelength calibration. In addition, we will be stepping the standard star across the chemically-etched and wire EDM slits to compare the stability of spectrophotometric data quality across the slits, as there is larger variation in slit width for the machined slits. From the lessons learned from testing this method of slit fabrication, we will bring the process to industry standards with multiple improvements. Improvements in the fabrication process in future work include using a buffered-oxide etchant prior to the potassium hydroxide etch to better control the consistency of etch time and slit widths, longer RIE etch times for the removal of remaining nitride, removal of the dicing tape by peeling instead of with acetone, and the addition of a metal backing substrate to stabilize the fragile slits. Additionally, we will characterize future slits using a scanning electron microscope for improved resolution over the optical microscope used in this paper.
 
To accommodate the specific needs of the APO user community, there is potential for fabrication and testing of multi-object reflective slits, for example multiple short slits. All that would be required from the user is a CAD file with the design of the slit mask. The lead time for these requests could be a week or two depending on the availability of tools at the WNF. Further testing will assess the feasibility of multiple short slits and spacing required to maintain integrity of the slit.

\section{Acknowledgements}

Special thank you to the engineers at the the Washington Nanofabrication Facility who taught this astronomer how to use nanofabrication equipment, in particular David Nguyen, Mark Morgan, Mark Brunson, Sarice Jones, Jean Nielsen, Darick Baker, and Duane Irish. Your patience and thoughtfulness with my endless questions are much appreciated! Thank you to Don Brownlee and Dave Joswiak for letting me use the Cosmic Dust Lab's microscope to image the slits. 

Part of this work was conducted at the Washington Nanofabrication Facility / Molecular Analysis Facility, a National Nanotechnology Coordinated Infrastructure (NNCI) site at the University of Washington, which is supported in part by funds from the National Science Foundation (awards NNCI-1542101, 1337840 and 0335765), the National Institutes of Health, the Molecular Engineering \& Sciences Institute, the Clean Energy Institute, the Washington Research Foundation, the M. J. Murdock Charitable Trust, Altatech, ClassOne Technology, GCE Market, Google and SPTS.

Part of this work was based on observations obtained with the Apache Point Observatory 3.5-meter telescope, which is owned and operated by the Astrophysical Research Consortium.

\bibliography{slits}   
\bibliographystyle{spiejour}   


\listoffigures
\listoftables

\end{spacing}
\end{document}